\documentclass[12pt,preprint]{aastex}
\received{ }
\revised{ }
\accepted{ }
\ccc{code}
\cpright{type}{year}
\usepackage{amsmath}
\usepackage{amssymb}
\usepackage{graphics}
\usepackage{graphicx}
\usepackage{natbib}

\shorttitle{North South Asymmetry in Photospheric B-flux}
\shortauthors{Shetye, Tripathi and Dikpati}

\begin{document}

\title{Observations and modelling of North-South asymmetries using a Flux Transport Dynamo}
\author{Juie Shetye}
\affil{Armagh Observatory, College Hill, Armagh City, Northern Ireland,UK,BT61 9DG}
\author{Durgesh Tripathi}
\affil{Inter-University Centre for Astronomy and Astrophysics, Post Bag - 4, Ganeshkhind, Pune 411007, India}
\and
\author{Mausumi Dikpati}
\affil{High Altitude Observatory, National Center for Atmospheric Research Boulder, Colorado USA 80307}

\begin{abstract}

The peculiar behaviour of the solar cycle 23 and its prolonged minima has been one of the most studied problems over the last few years. 
In the present paper, we study the asymmetries in active region magnetic flux in the northern and southern hemispheres during complete 
solar cycle 23 and rising phase of solar cycle 24. During the declining phase of solar cycle 23, we find that the magnetic flux in the 
southern hemisphere is about 10 times stronger than that in the northern 
hemisphere during the declining phase of the solar cycle 23 and during the rising
phase of cycle 24, however, this trend reversed. The magnetic flux
becomes about a factor of 4 stronger in the northern hemisphere to that of 
southern hemisphere. Additionally, we find that there was significant delay 
(about 5 months) in change of the polarity in the southern hemisphere in 
comparison with the northern hemisphere. These results provide us with hints
of how the toroidal fluxes have contributed to the solar dynamo during 
the prolonged minima in the solar cycle 23 and in the rising phase of the solar
cycle 24. Using a solar flux-transport dynamo model, we demonstrate that
persistently stronger sunspot cycles in one hemisphere could be caused by the
effect of greater inflows into active region belts in that hemisphere. 
Observations indicate that greater inflows are associated with stronger
activity. Some other change or difference in meridional circulation between
hemispheres could cause the weaker hemisphere to become the stronger one.

\end{abstract}

\newpage

\section{Introduction}

Based on the observations of Sunspots on the surface of the Sun and the relatively well-organised poloidal field, which 
changes polarity approximately every 11 year, it has been universally accepted that the solar magnetic cycle is a dynamo 
process involving the transformation of the polar field into the toroidal field and subsequent conversion of the toroidal field 
into the poloidal field of opposite polarity over the course of approximately 11 years \citep[see e.g.] [and its citations]
{Bab:1961}. The generation and propagation of large-scale magnetic fields via the dynamo mechanism is considered to be 
a two-step process. The first step involves shearing of the poloidal component of magnetic field by differential rotation, 
which gives rise to the azimuthally directed toroidal magnetic field. This toroidal field then gives rise to the formation of the
sunspots and the active regions (henceforth ARs). The second step is the formation of the poloidal component from the 
toroidal component, which occurs from the magnetic flux liberated by the growth and the decay of the sunspots, with the 
leading polarity flux moving towards the equator and the following polarity towards the pole. In some models, movement 
of the following polarity fields towards poles is due to the meridional circulation, as illustrated with both the kinematic 
dynamo and the flux-transport dynamo models \citep[see e.g.][and their citations]{WanSN:1991,ChoSD:1995}. 

The study of the solar cycle on long time scales indicates that the solar cycle is virtually symmetric between North and 
South hemispheres, in the sense that the average amplitudes, shapes and durations of cycles are very similar
\citep[see e.g.,][]{GoeC:2009}. However, there are individual cycles that are known to be stronger in one hemisphere 
than other. For example, just after the Maunder Minimum, almost all the sunspots were observed in southern hemisphere \citep[see e.g.][]{RibN:1993}. 
 Asymmetries between the two hemispheres have also been observed in various solar activity 
phenomena such as sunspot area, sunspot numbers, faculae, coronal structure, post-eruption arcades, coronal ionisation 
temperatures, polar field reversals as well as solar oscillations \citep[see e.g.,][and reference therein]{ChoCG:2013,SykR:2010,
GaoLS:2009,LiEtal:2009,TemEtal:2006,KnaSBa:2004,KnaSB:2005,TriBC:2004,AtaO:1996,OliB:1994,Zolo2010,Sval2013}, in addition to long 
term hemispheric asymmetries in solar activity in previous solar cycles \cite[see e.g.,][]{VizB:1989,CarOB:1993,Nort2010}. 

Further, the rise and fall of solar cycle 23 has been discussed by many authors; it has been found that the behaviour of 
solar cycle 23 is very peculiar for an odd-number cycle \citep{ChoCG:2013}. Cycle 23 showed a slow rise compared to
other odd-numbered cycles and was found to be  weak compared to other odd-numbered cycle 
\citep{LiEtal:2009, ChoCG:2013}. Additionally, it shows an unusual second peak during the declining phase 
\citep{LiEtal:2009, MisM:2012}.  Moreover, the temporal characteristics of cycle 23, such as sunspot number and sunspot 
area, are similar to the Gleissberg global minimum cycles 11, 13 and 14, which occurred between 1880 and 1930, as well 
as solar cycle 20 \citep{Kra:2012}. Analysis of polar field patterns indicate that polar field reversal was slower than the 
previous two cycles as discussed in \citep{Dik:2004}, which could have delayed the rise of solar cycle 24. The first two 
years of cycle 24, with low solar activity concentrated in the South, is similar to the cycle that immediately followed the 
Maunder Minimum \citep{Kra:2012}.

Our motivation for this research is to investigate solar cycle 23 and rise of solar cycle 24 by computing the AR fluxes 
that form the toroidal fluxes. We then concentrate on the solar cycle minimum and investigate the asymmetry in the 
hemispheres with the aim of addressing the issue related to the deep minimum observed in cycle 23. To support our observations, 
we have carried out dynamo simulations as mentioned in \citet{2013ApJ...779....4B}. We also investigated the role of 
meridional circulation combined with flux asymmetry for the solar cycle 23 by discussing different cases related to the
asymmetries found. The rest of the paper is organised as follows: in Section~\ref{obs} we present the  observations and 
data selection, followed by magnetic flux analysis and results in Section~\ref{res}. In Section~\ref{models} we discuss dynamo simulations 
and relate them to our observations. In the concluding  Section~\ref{summary}, we summarise the results and science.

\section{Observations and Data} \label{obs}

\begin{figure*}
\centering
\includegraphics[width=0.4\textwidth , height=0.4 \textheight, clip]{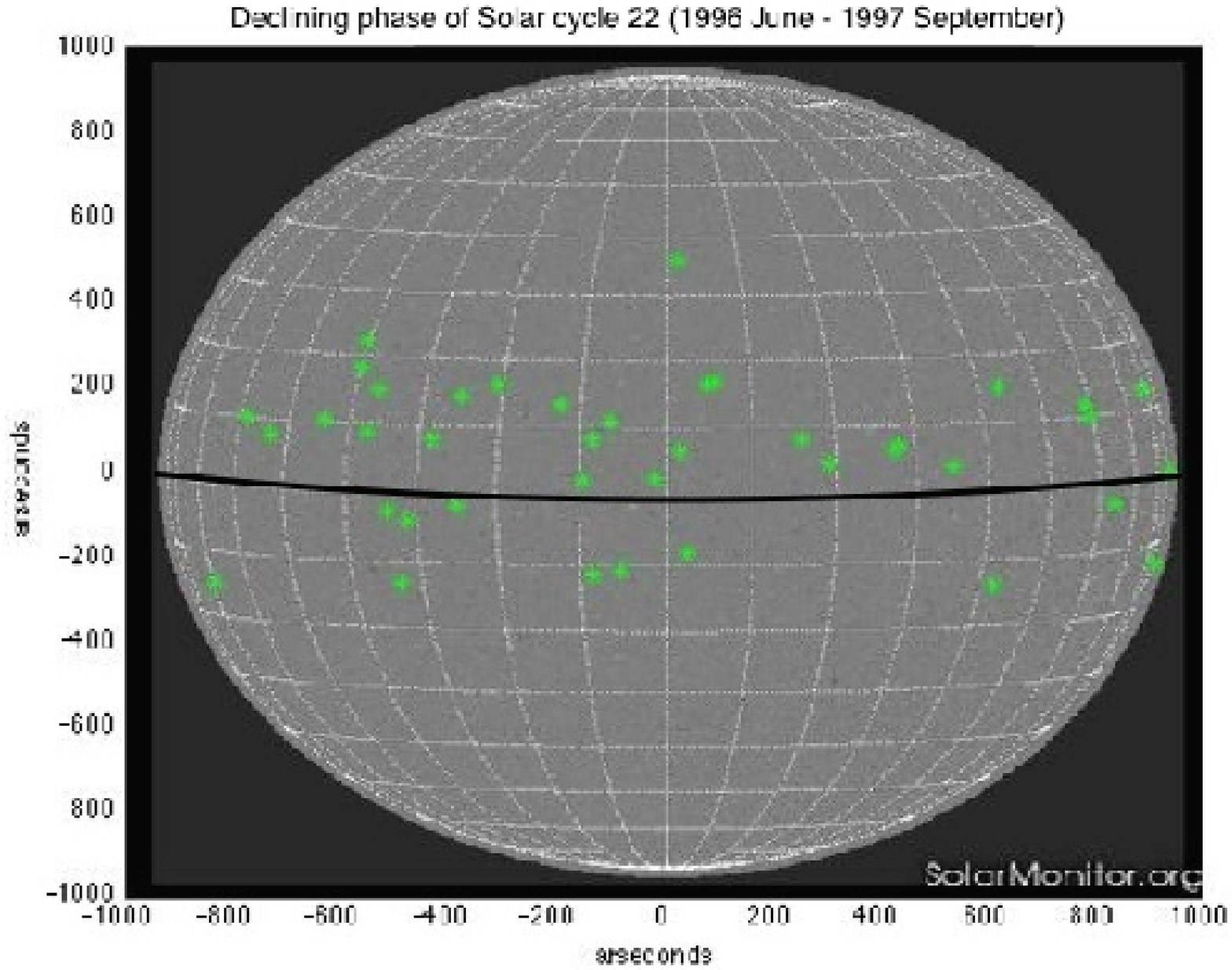}
\includegraphics[width=0.4\textwidth , height=0.4 \textheight, clip]{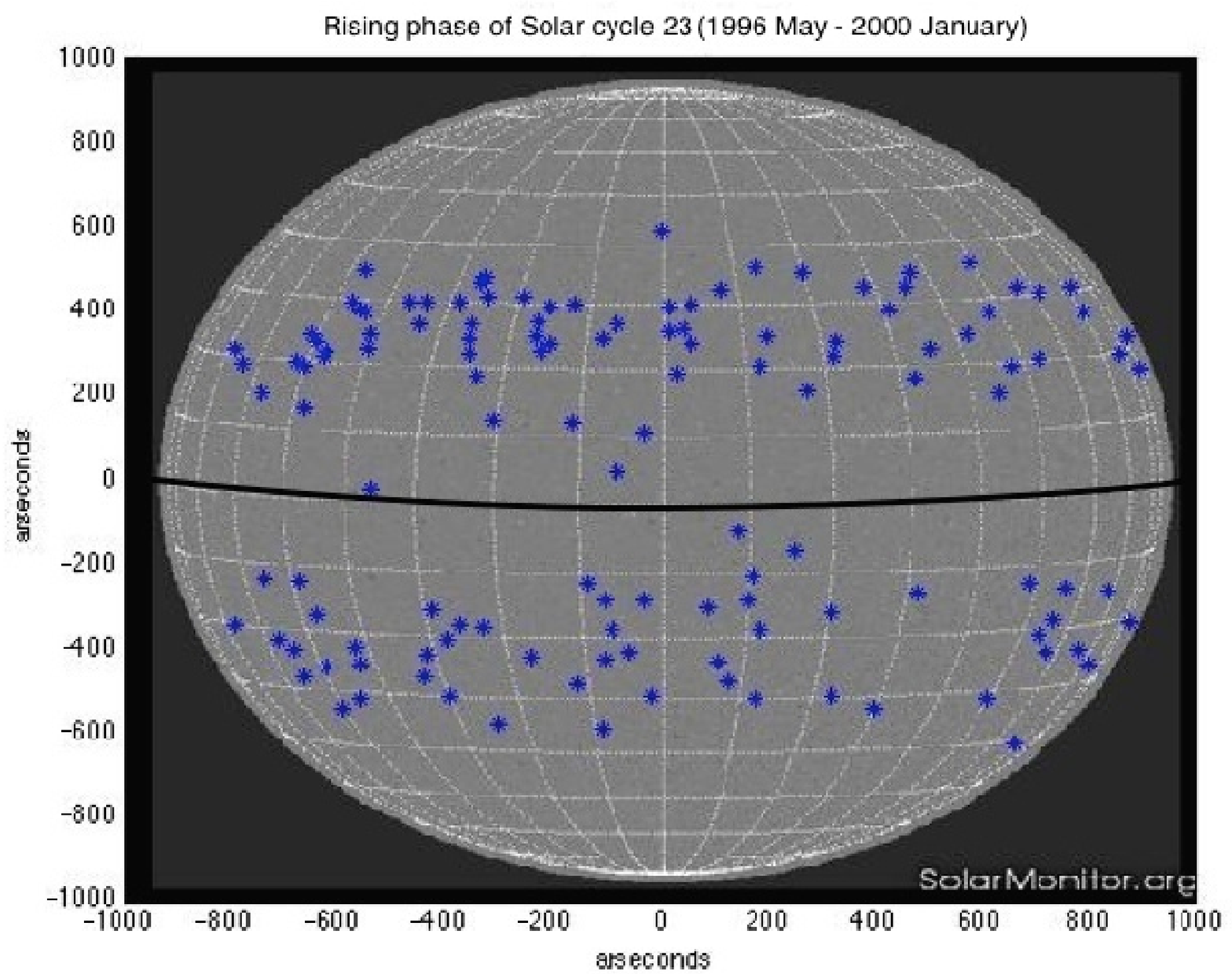}
\includegraphics[width=0.4\textwidth , height=0.4 \textheight, clip]{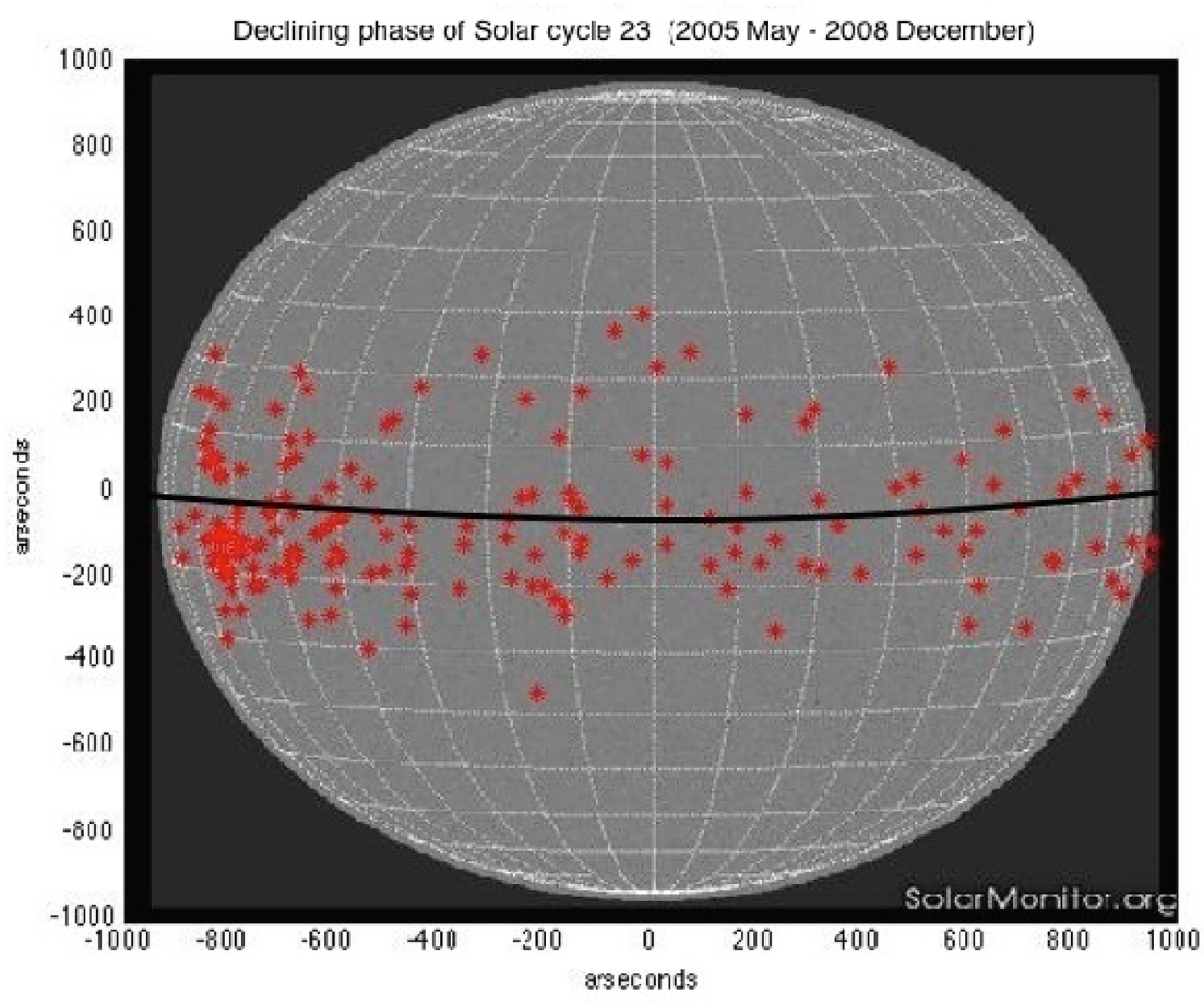}
\includegraphics[width=0.4\textwidth , height=0.4 \textheight, clip]{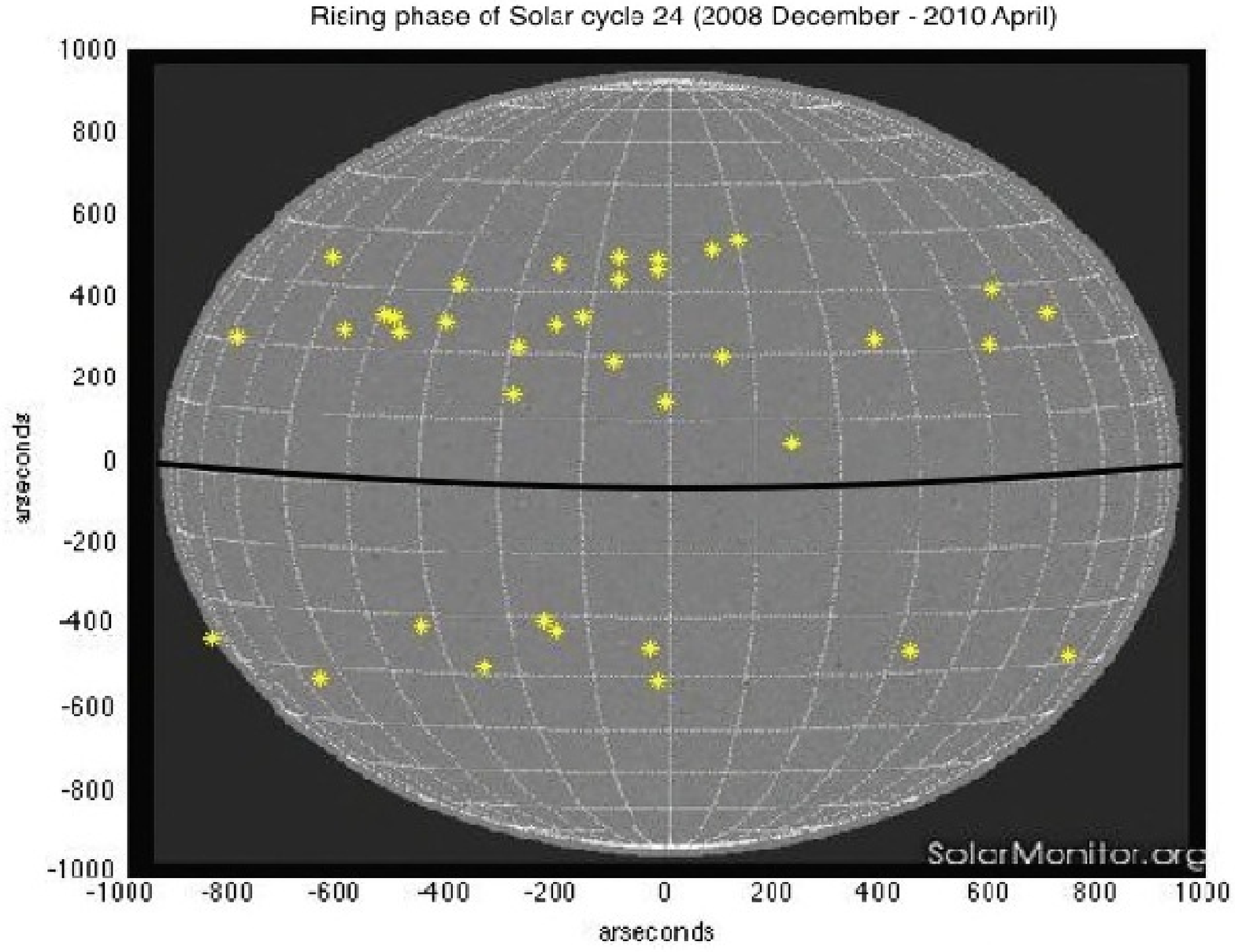}
\caption{Position of all the active regions formed between 2005 July 22 to 2010 April 12 during decline phase 
of the solar cycle 22 (top left), rise phase of the solar cycle 23 (top right), decline phase of the solar cycle 23 
(bottom left) and rise phase of cycle 24 (bottom right). The location of active region is based on its nomenclature 
from NOAA database.
\label{decliningcycle23}}
\end{figure*}

We calculate the line-of-sight (LOS) component of the magnetic field (hereafter B$_{\bot}$) using the magnetograms 
recorded by the Michelson Doppler Imager \citep[MDI;][]{SchEtal:1995} on board the Solar Heliospheric Observatory 
(SoHO). MDI is an instrument used to observe signs and strength of the line-of-sight component of the photospheric 
magnetic field. MDI images the Sun using a 1024 $\times$ 1024 CCD camera and acquires one full-disk line-of-sight 
magnetic field each 96 minutes (five minutes averaged 96-min cadence), among other observing sequences, which is 
free from atmospheric noise. A full-disk magnetogram of the Sun has a resolution of $\sim$ 4\arcsec\,
(2\arcsec $\times$ 2\arcsec \, per pixel) and a field of view of 34\arcmin\, $\times$ 34\arcmin. Pre-flight per pixel error 
in the flux was estimated at 20 Gauss (20 Mx cm$^{-2}$) \citep{SchEtal:1995}, which was found to be 14 Gauss  
in flight as was reported by \cite{Hag:2001}.

In the present work, we have used MDI magnetograms to compute the daily magnetic flux of ARs observed by \textit{solarmonitor.org} on the solar disk from 1996 May 06 to 2010 April 12 (approximately 5100 days) that covers the final stages of the solar cycle 22, the complete cycle 23 
and rising phase of the cycle 24. During this period, we have manually monitored evolution of 1948 ARs, which include 286 AR nests 
and 6 AR evolutions, where dispersion stage of ARs was observed persistent over multiple revolutions. We preferred using a manual approach than an automated one discussed in 
\citet{Zhang2010} and \citet{Sten2012},  so that we could eliminate multiple counting of the same AR due to solar rotation. 

\begin{figure}[h!]
\centering
\includegraphics[width=1.1\textwidth]{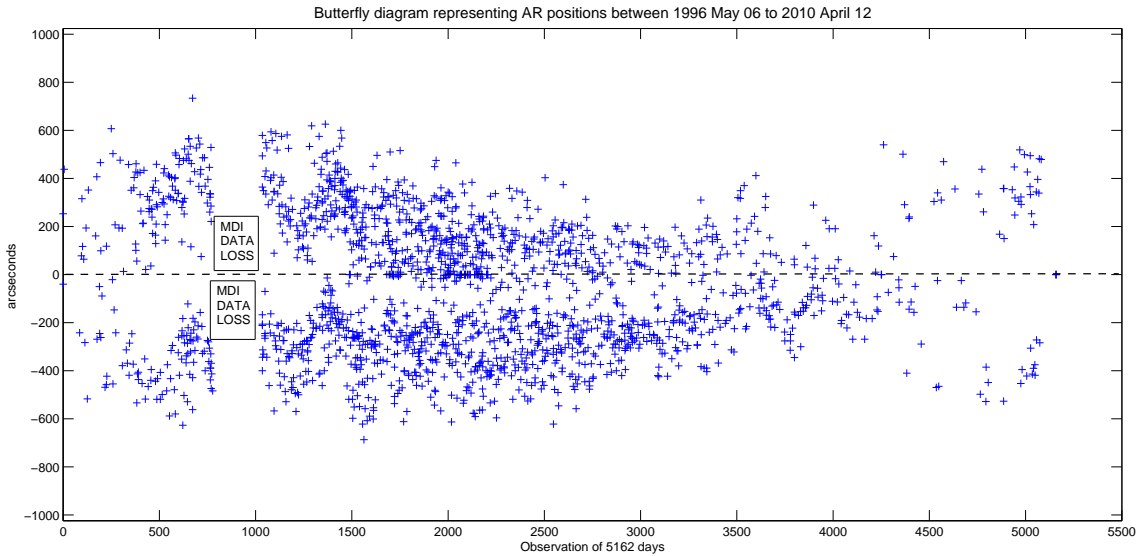}
\caption{A butterfly diagram presenting the position of AR found during 1996 May 06 to 2010 March 10. }
\label{butterfly_o}
\end{figure}

Fig.~\ref{decliningcycle23} shows the location of ARs on the solar disk during declining phase of the cycle 22 (top left), the rising phase
of the solar cycle 23 (top right), the declining phase of the solar cycle 23 (bottom left) and the rising phase of the solar cycle
24 (bottom right). The two hemisphere are separated by black line representing the solar equator. We further represent our 
observations in a butterfly diagram in Fig.~\ref{butterfly_o} that shows the location of AR observed and the time. Here, in the case of 
AR nests we have noted the position and date of AR that was first detected in \textit{solarmonitor.org}.
 We then concentrate our analysis on the 238 ARs observed on the solar disk from 2005~July~22 to 
2010~April~12, which covers the declining phase of cycle 23 and the rising phase of cycle 24. 

To refine the dataset, we ignore those ARs which dispersed in less than two days, since they generally lacked sunspots.
We also exclude ephemeral ARs their fluxes are discussed in \citet{Hag:2001} and \citet{ HagST:2003}. The MDI 
magnetogram shows noise of 0.02\arcsec on the limbs, with additional noise on the right limb due to wavelength changes 
in the Michelson filters $\Theta$ \citep{WenSKF:2004}. Hence we ignore all ARs born on the right limb at longitudes  
$\gtrsim$ 60\degr. For ARs born on the right limb with an angle $\gtrsim$ 60\degr the angle between LOS magnetic field 
and perpendicular magnetic field (${\bot}$) tends to zero \citep{Hag:2001,HagST:2003}. Hence we have omitted all 
readings beyond $\gtrsim$ 60\degr. 
 
After choosing the ARs, we select a box containing isolated region/s and calculate the highest value of B$_{\bot}$ viz. 
B$_{\bot max}$ in each box. Thereafter we separate the negative and positive polarities of Active Regions, using contours 
with levels at B$_{\bot}$ = 0, B$_{\bot}$ \textless \, -14 Gauss and B$_{\bot}$ \textgreater \, 14 Gauss. Then we define a contour 
at 99 \% of the calculated B $_{\bot max}$ values to eliminate the magnetic fields outside the Active Regions. Finally, we 
use the equation~\eqref{eq1} to calculate the magnetic fluxes ($\Phi$) of each regions separately, 

\begin{equation}  \label{eq1}
 \Phi  = \int B \cdot ds
\end{equation}

Since the AR magnetic field is bipolar, the net unbalanced magnetic flux in an AR should be nearly zero. To check if our 
observations were accurate we noted the net unbalanced flux; if it has a value other than zero, then we return to the 
first step again and reselect the boxes. This parameter also meant that we consider AR nest as single large AR that occurred 
mostly during maximum of solar cycle 23. During the minima we found two AR nests and were able to resolve them into individual ARs. 
In order to remove the error due to line-of-sight (LOS) effects, we calculate the 
angle $\Theta $ of each pixel from the disk center following method described in \citet{Hag:2001} using:

\begin{equation}  \label{eq:eq2}
\Theta = \sin^{-1} [ \frac{\sqrt{ \left( \begin{array}{c} {x_{\bot}}^{2}
\end{array} \right)  +
\left( \begin{array}{c} {y_{\bot}}^{2} \end{array} \right)  }} {R} ]
\end{equation}

(where x$_{\bot}$ , y$_{\bot}$ are radial co-ordinates of each pixel within the active region on the solar disk and R is the 
radius of Sun in pixels).

To calculate the LOS magnetic flux of ARs, we have used the relation that LOS magnetic flux is related to perpendicular 
component of magnetic flux  using the formula
 
\begin{equation} 
 \Phi_{LOS}= \Phi_{\bot} \cdot \cos\Theta.
\end{equation} 

\section{Analysis and results related to the deep minima}\label{res}

\begin{figure}[h!]
\centering
\includegraphics[width=1.0\textwidth]{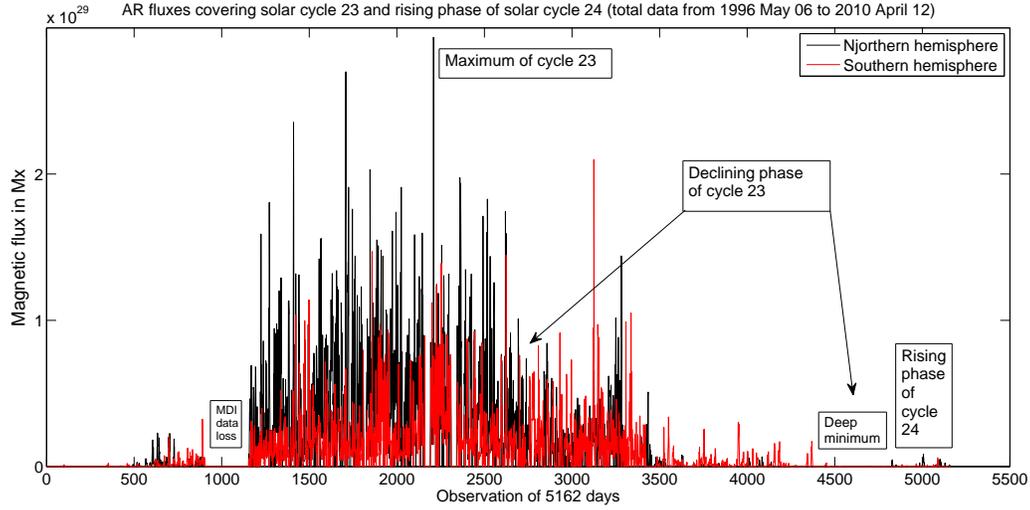}
\caption{Comparison of AR flux behaviour between the northern and the southern hemispheres from the 1996 May 06 to 2010 March 10.}
\label{completedataset}
\end{figure}

\begin{figure}[htbp]
\centering
\includegraphics[width=1.1\textwidth, height=0.3\textheight ]{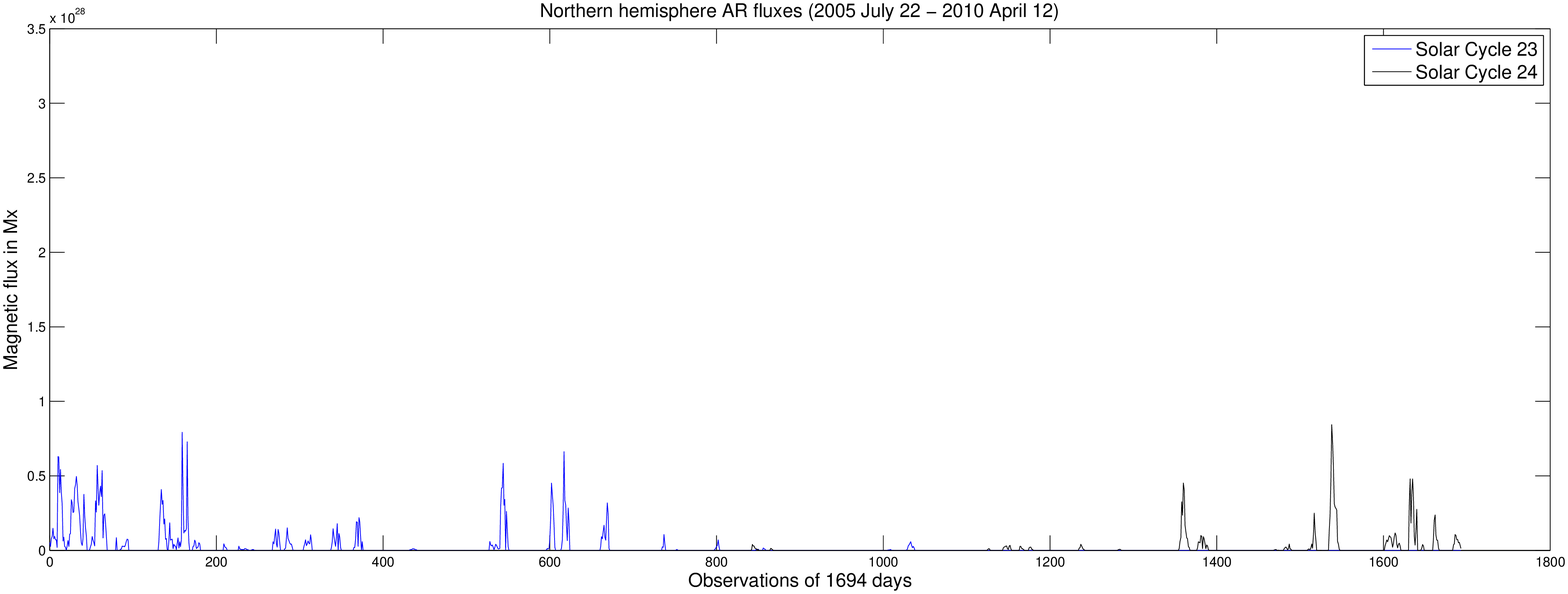}
\includegraphics[width=1.1\textwidth, height=0.3\textheight ]{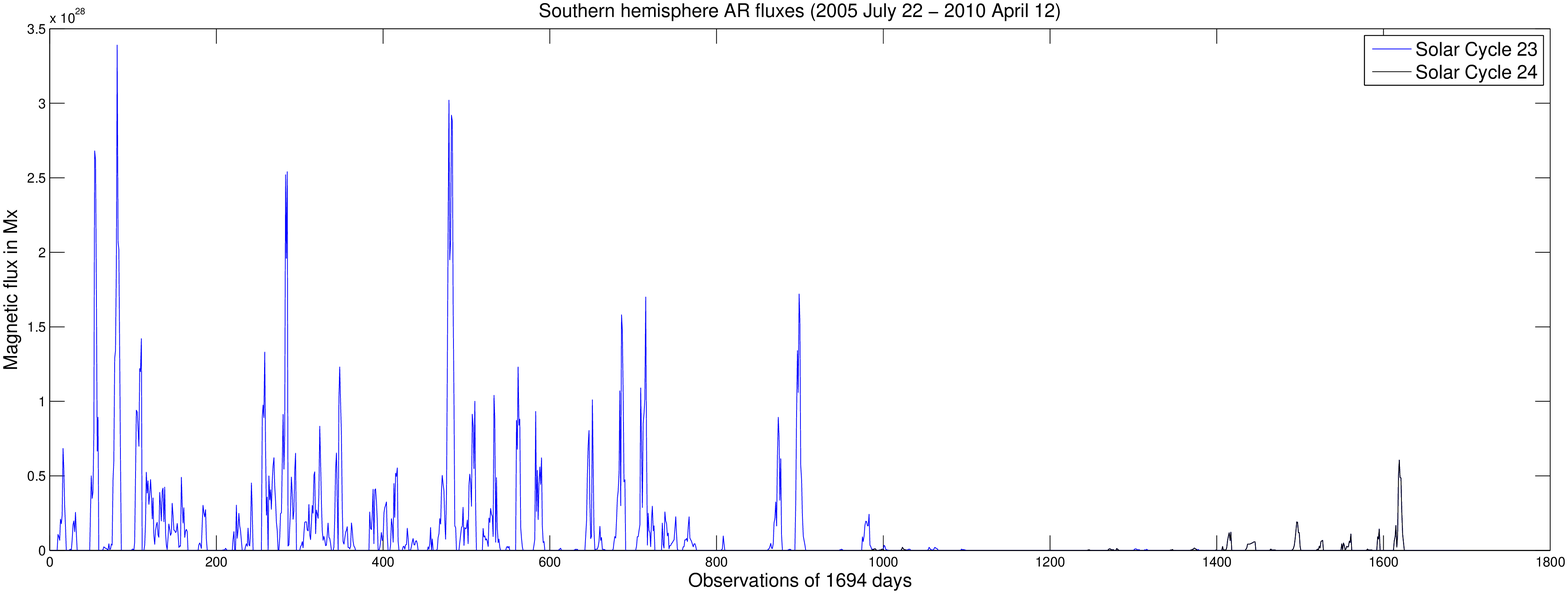}
\caption{AR flux behaviour during the deep minimum, the upper panel shows flux behaviour in the Northern hemisphere and the lower panel 
shows AR flux behaviour in the Southern hemisphere. The graphs clearly indicate the difference in strengths of magnetic fluxes in both the 
cycles\label{Fluxescompared}.} 
\end{figure}

In order to understand magnetic flux evolution, we produce magnetic flux vs time graphs as shown in Fig.~\ref{completedataset} 
and Fig.~\ref{Fluxescompared}. We summarise the calculated fluxes from the complete database in the Fig.~\ref{completedataset}. We 
represent fluxes from NOAA 07961 as 0 (1996 May 06) on the x-axis as the beginning of the dataset and proceed with computing daily 
fluxes. Since we intended to study both hemispheres independently, we have represented AR fluxes in the northern hemisphere by black 
solid lines and southern hemisphere by red solid lines. It can be easily noticed from Fig.~\ref{completedataset} that the solar cycle 23 
was significantly asymmetric in terms of dominating hemispheres.

\cite{Sten2012} and \cite{Zhang2010}, have carried detailed analysis of active regions observed by MDI  with their respective techniques. 
\citet{Sten2012}, performed critical analysis of dataset using automated techniques and studied various properties of these bipolar regions including
their orientation  including the tilt angle variation. \cite{Zhang2010}, on the other hand studied the basic physical parameters including the magnetic flux of
individual active regions, their distribution with respect to size as well as magnetic flux. They found the solar cycle between 1996 and 2008 very asymmetric. In terms of number of ARs, they found 938 ARs in the South and 792 in the North. We found similar results, we found 1071 AR in the South and 872 in the North, which makes the asymmetry approximately 12 \%. \cite{Zhang2010} also studied the asymmetry in the magnetic flux and found that the Southern hemisphere was stronger than the Northern hemisphere during the declining phase of the solar cycle 23, we found a similar asymmetry where we found 10 times more flux emergence in the Southern hemisphere. These results are also in agreement with those obtained by 
\citep{LiEtal:2009} and \citep{ChoCG:2013}, based on sunspot  observations.

Fig.~\ref{completedataset} suggest that the behaviour of deep minima may be related to AR fluxes related to the later 
part of the declining phase of the cycle 23. Thus we concentrate our analysis on the latter part of the cycle 23. In order to 
study the flux behaviour therein, we selected final 1694 days from Fig.~\ref{completedataset} (i.e. data between 
3468$^{th}$ day to 5162$^{th}$ day) and represented in the graphs in Fig.~\ref{Fluxescompared}. Here, we represent 
our observations with NOAA AR 10791 (northern hemisphere, observed on 2005 July 22 and represented by 0 on 
the x-axis in the top panel of Fig.~\ref{Fluxescompared}). In the South we began with NOAA AR 10794 (southern 
hemisphere, observed on 2005 August 01 ), both occurring roughly mid-way during the declining phase of cycle 23. We continue until 
the dispersion of NOAA AR 11060 (in the northern hemisphere on 2010 April 12 represented by 1694 on the x-axis in 
Fig.~\ref{Fluxescompared}) in the North. In the plots, blue line indicates the magnetic flux behaviour during the 
declining phase of cycle 23 and the black line indicates the magnetic flux behaviour during the rising phase of cycle 24.

Fig.~\ref{Fluxescompared} clearly indicates that photospheric magnetic flux during the final four years of solar cycle 23 was dominant in the southern hemisphere, producing a profound north-south asymmetry in terms of AR numbers and magnetic flux. During this period we found 121 ARs in the Southern hemisphere as compared to 60 ARs in the Northern hemisphere. This 
asymmetry becomes even more pronounced as the cycle progresses. The AR fluxes in the southern hemisphere were 
approximately 10 times stronger than those in the northern hemisphere. But this behaviour changed completely during 
the rising phase of cycle 24, for which the strength of fluxes in the North is 4 times that of the South. This was observed with 36 ARs emerging in the North compared to 23 in the South. Proceeding towards 
cycle 24, we find that (see Fig.~\ref{Fluxescompared}) the new cycle began in the North on the 875 $^{th}$ day, which 
is 2007 December 13 with the emergence of opposite polarity sunspot was observed. In the South the opposite polarity sunspot was observed on 1013$^{th}$ day, which is 2008 May 3. This is 143 days (about 5 months) after the reversal of polarities in northern 
hemisphere. Moreover, Fig.~\ref{Fluxescompared} also clearly shows that the change of polarity in the northern hemisphere 
occurred smoothly and quickly, with a mixture of ARs from both cycles for a period of 200 days, whereas in the southern 
hemisphere, the new emerging flux showed a delay.  

\section{North-South asymmetries from dynamo action}\label{models}

In order to understand how the solar dynamo works and to investigate the reason behind our 
observational results concerning the asymmetry between hemispheres, and to understand the 
effect of this asymmetry on the deep minima, we have carried out sets of dynamo simulations.
\citet{2013ApJ...779....4B} have shown that differences in the form and amplitude of meridional circulation 
between North and South hemispheres can cause significant differences in the poloidal, polar and toroidal fields produced 
there. The longer the meridional circulation differences persisted, the larger the differences became. 
\citet{2013ApJ...779....4B} focussed on global changes in meridional circulation, including amplitude changes of the whole 
circulation, differences between one and two cells in either latitude or depth. The meridional motions of sunspots, pores \citep[and 
their citations]{1990GApFD..55..241R} and other magnetic features \citep{1994SoPh..149..417K,1999ApJ...527..967M} are related to 
generation of inflows. Observations by \citet{2005LRSP....2....6G} 
show that there are also significant meridional circulation patterns in the Sun that are not global, including one that is
associated with ARs themselves. In particular, there are inflows into ARs from lower and higher latitudes 
that can be as large as 50~m~s$^{-1}$. When averaged in longitude, these inflows can create a meridional circulation 
signal of 5~m~s$^{-1}$ or more. With more active regions in one hemisphere compared to the other, the average inflow 
should also be larger in the more active hemisphere. In addition, since ARs are the source of surface poloidal flux that 
migrates toward the poles and causes polar field reversals, the effect of the meridional circulation from the inflows may in 
fact be larger than represented by the full longitude average. It has also been shown that the inflows may play an important
role in the generation of poloidal field during the final stages of the solar cycle \citep{Came2012, Jiang2010}. Therefore, we need to 
assess the role of active region inflows and their differences between North and South hemispheres to see how much difference in solar 
cycles they can produce. 

We have carried out flux-transport dynamo simulations using the same model as used in \citet{2013ApJ...779....4B}, 
in order to see the role of active region inflows. For the sake of completeness, we briefly
repeat the set-up of the simulation runs in the following subsection (\S4.1), which
describes the dynamo equations, mathematical forms of the dynamo ingredients and 
boundary and initial conditions. The subsection \S4.2 presents the detailed formulation 
of the treatment of inflow cells, and \S4.3 the consequences of the inflow cells.

\subsection{Dynamo simulation set-up}\label{ingredients}

Our starting point is the set-up of \citet{2013ApJ...779....4B}. We write the dynamo 
equations as:

$$\frac{\partial A}{\partial t} + \frac{1}{r\sin\theta}({\bf u} .\nabla)
(r\sin\theta A) = \eta({\nabla}^2 - \frac{1}{r^2 \sin^2 \theta}) A
+\left[S_{\rm BL}(r,\theta) + S_{\rm tac} (r,\theta)\right]$$
$$ \times {\left[1+{\left(\frac{B_{\phi}|_{ov}(\theta,t)}
{{B_0}}\right)}^2 \right]}^{-1} \thinspace B_{\phi}|_{ov}(\theta,t),
\eqno(4a)$$
$$\frac{\partial B_{\phi}} {\partial t} + \frac{1}{r} \left[\frac{\partial}
{\partial r}(r u_r B_{\phi}) + \frac{\partial}{\partial \theta}(u_{\theta}
B_{\phi}) \right]= r\sin\theta ({\bf B}_p .\nabla)\Omega
-{\bf{\hat e}}_{\phi}\thinspace . \thinspace \left[
\nabla\eta\times\nabla\times B_{\phi}{\bf{\hat e}}_{\phi}\right]
$$
$$+\eta({\nabla}^2
-\frac{1}{r^2 \sin^2 \theta})B_{\phi},\eqno(4b)$$
in which $A(r,\theta,t)$ denotes the vector potential for the poloidal field, 
$B_{\phi}(r,\theta,t)$ the toroidal field, $u_r (r,\theta),\, u_{\theta}(r,\theta)$ 
the meridional flow components, $\Omega(r,\theta)$ the differential rotation, 
$\eta(r)$ the depth-dependent magnetic diffusivity, $S_{\rm BL}(r,\theta)$ the 
Babcock-Leighton type surface poloidal source, $S_{\rm tac} (r,\theta)$ the 
tachocline $\alpha$-effect and $B_0$ the quenching field strength, which we set
to 10 kGauss in this calculation. 

We use the following expressions respectively for Babcock-Leighton
surface source and tachocline $\alpha$-effect:

$$ S_{\rm BL}(r,\theta)
=\frac{s_1}{4}\left[1+{\rm erf}\left(\frac{r-r_1}{d_1}\right)\right]
\left[1-{\rm erf}\left(\frac{r-r_2}{d_2}\right)
\right]\, {2\, \sin\theta \cos\theta
}. \quad \eqno(5)$$
For $0 \le \theta \le \pi/2$,
$$ S_{\rm tachocline}(r,\theta) =
\frac{s_2}{4}
\left[1+{\rm erf}\left(\frac{r-r_3}{d_3}\right)\right]
\left[1-{\rm erf}\left(\frac{r-r_4}{d_4}\right) \right].
\frac{\sin\left(6(\theta-\frac{\pi}{2})\right) e^{-\gamma_2 \thinspace
{(\theta-\frac{\pi}{4})}^2}} {[e^{\gamma_3(\theta- \pi/3)} +1]},
\eqno(6a)$$
and, for $\pi/2 \le \theta \le \pi$,
$$ S_{\rm tachocline}(r,\theta) =
\frac{s_2}{4}
\left[1+{\rm erf}\left(\frac{r-r_3}{d_3}\right)\right]
\left[1-{\rm erf}\left(\frac{r-r_4}{d_4}\right) \right].
\frac{\sin\left(6(\theta-\frac{\pi}{2})\right) e^{-\gamma_2 \thinspace
{(\theta-\frac{3\pi} {4})}^2}} {[e^{\gamma_3 (2\pi/3-\theta)} +1]}.
\eqno(6b)$$
The parameter values used in (5), (6a) and (6b) are: $s_1=2.0 \,{\rm m}
{\rm s}^{-1}$, $s_2=0.5 \,{\rm m}{\rm s}^{-1}$, $r_1=0.95R_{\odot}$,
$r_2=0.987R_{\odot}$, $r_3=0.705R_{\odot}$, $r_4=0.725R_{\odot}$, 
$d_1=d_2=d_3=d_4=0.0125R_{\odot}$,
$\gamma_2=70.0$, $\gamma_3=40.0$. Note that the values of $s_1$ and
$s_2$ determine the amplitude of the Babcock-Leighton poloidal source
term and the tachocline $\alpha$-effect respectively, but the maximum
amplitudes of $S_{\rm BL}$ and $S_{\rm tachocline}$  are not exactly
$2 \,{\rm m}{\rm s}^{-1}$ and $50\,{\rm cm}{\rm s}^{-1}$, but instead
$\sim 1.93 \,{\rm m}{\rm s}^{-1}$ and $\sim 37\,{\rm cm}{\rm s}^{-1}$
respectively, for the parameter choices given above. This happens due
to the modulation of error functions used in expressions (5), (6a)
and (6b).

The diffusivity profile is given by (7) (for more details, see 
\citet{2002ApJ...575L..41D}.
$$\eta(r)=\eta_{\rm core}+\frac{\eta_{\rm T}}{2}\left[1+{\rm erf}\left(
\frac{r-r_5}{d_5}\right)\right]+\frac{\eta_{\rm super}}{2}\left[1+{\rm erf}
\left(\frac{r-r_6}{d_6}\right)\right], \quad\eqno(7)$$
in which, $\eta_{\rm core}=10^9 \,{\rm cm}^2 {\rm s}^{-1}$,
$\eta_{\rm T}=7\times 10^{10} \,{\rm cm}^2 {\rm s}^{-1}$,
$\eta_{\rm super}=3 \times 10^{12} \,{\rm cm}^2 {\rm s}^{-1}$, 
$r_5=0.7R_{\odot}$, $r_6=0.96R_{\odot}$, $d_5=0.00625R_{\odot}$, 
$d_6=0.025R_{\odot}$. These choices
make this profile possess a supergranular type diffusivity value
($\eta_{\rm super}$) in a thin layer at the surface, which drops to a
turbulent diffusivity value ($\eta_{\rm T}$) in the bulk of the convection
zone, and at the base of the convection zone the diffusivity drops quite
sharply to a much lower value ($\eta_{\rm core}$) to mimic the molecular
diffusivity. 

The stream function $\psi$ for the steady part of the meridional
circulation is given by:

$${\psi}r\sin\theta={\psi}_0 \frac{(\theta-{\theta}_0)}{(\theta+\theta_0)}
\sin\left[k \frac{\pi(r-R_b)}
{(R_{\odot}-R_b)}\right] \left(1-e^{-{\beta}_1 r{\theta}^{\epsilon}}\right)
\left(1-e^{{\beta}_2 r(\theta-\pi/2)}\right)
e^{-{((r-r_0)/\Gamma)}^2}. \quad\eqno(8)$$

The streamline flow can be obtained in the North hemisphere by plotting the 
contours of $\psi\,r\,\sin\theta$. The streamlines in the South hemisphere can be 
obtained by implementing mirror symmetry about the equator. This parameter values 
for this stream function are:
\noindent $k=1$, $R_b=0.69 R$, $\beta_1
=0.1/(1.09 \times 10^{10}) \, {\rm cm}^{-1}$,
$\beta_2=0.3/(1.09 \times 10^{10}) \,{\rm cm}^{-1}$,
$\epsilon=2.00000001$, $r_0=(R_{\odot}-R_b)/5$, $\Gamma=3 \times 1.09 \times
10^{10}\,{\rm cm}$ and $\theta_0=0$. This choice of the set of
parameter values produce a flow pattern that peaks at $24^{\circ}$
latitude.

In order to perform simulations in non-dimensional units, we use
$1.09 \times 10^{10} \,{\rm cm}$ as the dimensionless length and
$1.1\times 10^8 \,{\rm s}$ as the dimensionless time. These choices
respectively come from setting the dynamo wavenumber, $k_D=9.2
\times 10^{-11} \, {\rm cm}^{-1}$, as the dimensionless length, and
the dynamo frequency, $\nu=9.1 \times 10^{-9} \, {\rm s}^{-1}$, as
the dimensionless time, which means that the dynamo wavelength ($2\pi
\times 1.09 \times 10^{10} \, {\rm cm}$) is $2\pi$ and the mean dynamo
cycle period (22 years) is $2\pi$ in our dimensionless units. Thus,
in non-dimensional units, the parameters that define the meridional
circulation given in the expression (5) are: $R_b=4.41$, $\beta_1=0.1$,
$\beta_2=0.3$, $\epsilon=2.00000001$, $r_0=(R_{\odot}-R_b)/5$, $\Gamma=3$ and
$\theta_0=0$. The latitude of the peak flow can be varied by changing
$\beta_1$ and $\beta_2$; for example, changing $\beta_1$ from 0.1 to 0.8
and $\beta_2$ from 0.3 to 0.1, a flow pattern can be constructed that
peaks at $50^{\circ}$, but for the present study, we fix the latitude
of the peak flow at $24^{\circ}$.

Considering an adiabatically stratified solar convection zone, we
take the density profile as,
$$\rho(r) = \rho_b {\left[ (R_{\odot}/r)-1 \right]}^m, \quad\eqno(9)$$
in which $m=1.5$. However, in order to avoid density vanishing at
$r=R_{\odot}$, which would cause an unphysical infinite flow at the surface,
we use $\rho(r) = \rho_b {\left[ (R_{\odot}/r)-0.97 \right]}^m$ in
our simulations. Using the constraint of mass-conservation,  
the velocity components ($v_r, \, v_{\theta}$) can be computed from
$$u_r=\frac{1}{\rho r^2 \sin\theta}\frac{\partial}{\partial \theta} (\psi r
\sin\theta), \quad\eqno(10a)$$
$$u_{\theta}=-\frac{1}{\rho r \sin\theta}\frac{\partial}{\partial r} (\psi
r \sin\theta). \quad\eqno(10b)$$
The peak flow speed is determined by a suitable choice of $\psi_0/\rho_b$.
We use a peak flow speed of $14 \,{\rm m}{\rm s}^{-1}$ in all simulations.

\subsection{Formulation of inflow cells}\label{inflow}

In order to include inflow cells into the steady meridional circulation
pattern, described in \S4.1, we incorporate a time-dependent stream function
($\Psi_{\rm inflow}$), which is prescribed as follows:

$$\Psi_{\rm inflow} = {\psi}_{0_{\rm inflow}}\sin\left[\frac{\pi(r-r_{\rm inflow})} 
{(R_{\odot}-r_{\rm inflow})}\right] \sin\left[\frac{2\pi(\theta - 
\theta_{\rm high})} {(\theta_{\rm low})}\right]. \quad\eqno(11)$$

\noindent In expression (11), ${\psi}_{0_{\rm inflow}}$ determines the velocity
amplitude of the inflow cells, $r_{\rm inflow}$ determines how deep down the 
inflow cells extend from the surface, $\theta_{\rm high}$ and 
$\theta_{\rm low}$ determine their extent in $\theta$ (colatitude) coordinate,
$\theta_{\rm high}$ denoting the cell-boundary at the poleward side and 
$\theta_{\rm low}$ the equatorward side. Since the inflow cells are normally 
associated with active regions, their $\theta$ locations have to be function 
of time. We implement the time-dependence in the $\theta$ coordinate of
the inflow cells in accordance with the migration of latitude-zone of sunspots.
Thus we prescribe $\theta_{\rm high}$ and $\theta_{\rm low}$ as follows:
$$\theta_{\rm high}=\theta_{\rm center} - \pi/18, \quad\eqno(12a) $$  
$$\theta_{\rm low}=\theta_{\rm center} + \pi/18, \quad\eqno(12b) $$  
$$\theta_{\rm center}=\theta_{\rm center_{\rm initial}} + \delta t
\frac{\pi/6} {\tau}. \quad\eqno(12c) $$  
Here $\theta_{\rm center}$ is the center of the inflow cells, 
$\theta_{\rm center_{\rm initial}}$ is the starting location of the
center of the inflow cells and $\frac{\pi/6}{\tau}$ is the migration
speed of the center of the inflow cells. Note that the $\theta$-extent of 
each of the pair of inflow cells is $10^{\circ}$. So in order to make the
inflow cells migrate from $\sim 50^{\circ}$ latitude to the equator, we
have to make their center migrate from $\sim 40^{\circ}$ latitude to
$\sim 10^{\circ}$ latitude. Here $\tau$ is approximately one sunspot 
cycle period (i.e. half of a magnetic cycle period), and $\delta t$ is
the time-step for dynamo field evolution. For simplicity, 
we assume in this calculation that their extent in depth remains the same. 
We take $r_{\rm inflow}=0.9 \_ R_{\odot}$.

\begin{figure}[hbt]
\centering
\includegraphics[width=0.8\textwidth]{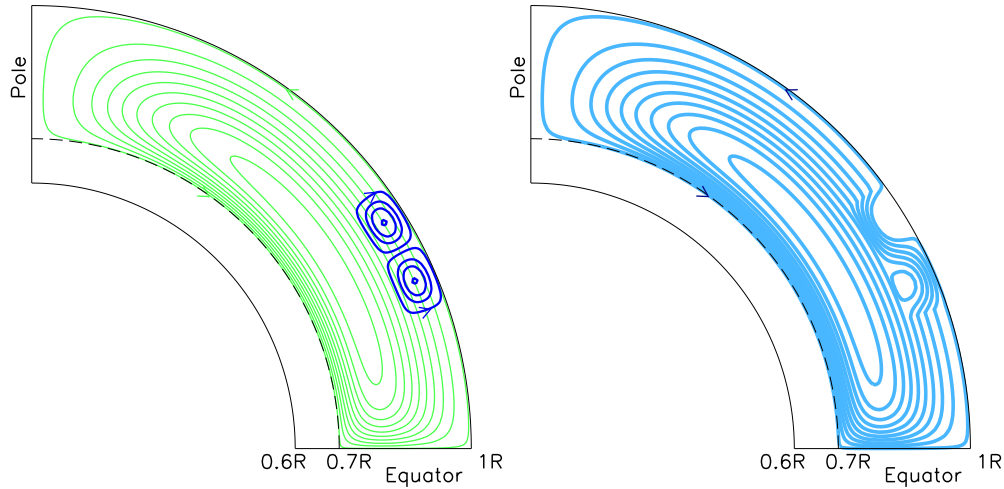}
\caption{Left image: Green streamlines show mean meridional circulation and on top of green streamlines inflow cells 
are separately plotted in blue. Right image shows the total flow pattern due to mean flow and inflow cells.}
\label{construct-inflowcell}
\end{figure}

Fig.~ \ref{construct-inflowcell} shows the prescribed form of the inflow 
circulation cells (see expressions 11, 12a-c) we have included in the 
dynamo model. In left panel we have superimposed the inflow circulation 
streamlines on the single celled global meridional circulation. As mentioned
earlier, in our calculations we have allowed the inflow circulation to 
reach to a depth of $0.9$ $R\sun$ and to latitudes of $10^{\circ}$ poleward 
and equator-ward of the active region latitude. The right panel in Fig.~\ref{construct-inflowcell} 
shows the total streamlines for a case for which the peak global circulation 
is 14~m~s$^{-1}$ and the peak inflow is 15~m~s$^{-1}$.

\subsection{Effect of north-south asymmetry in inflow cells}\label{inflow}

\begin{figure}[hbtp]
\centering
\includegraphics[width=0.8\textwidth]{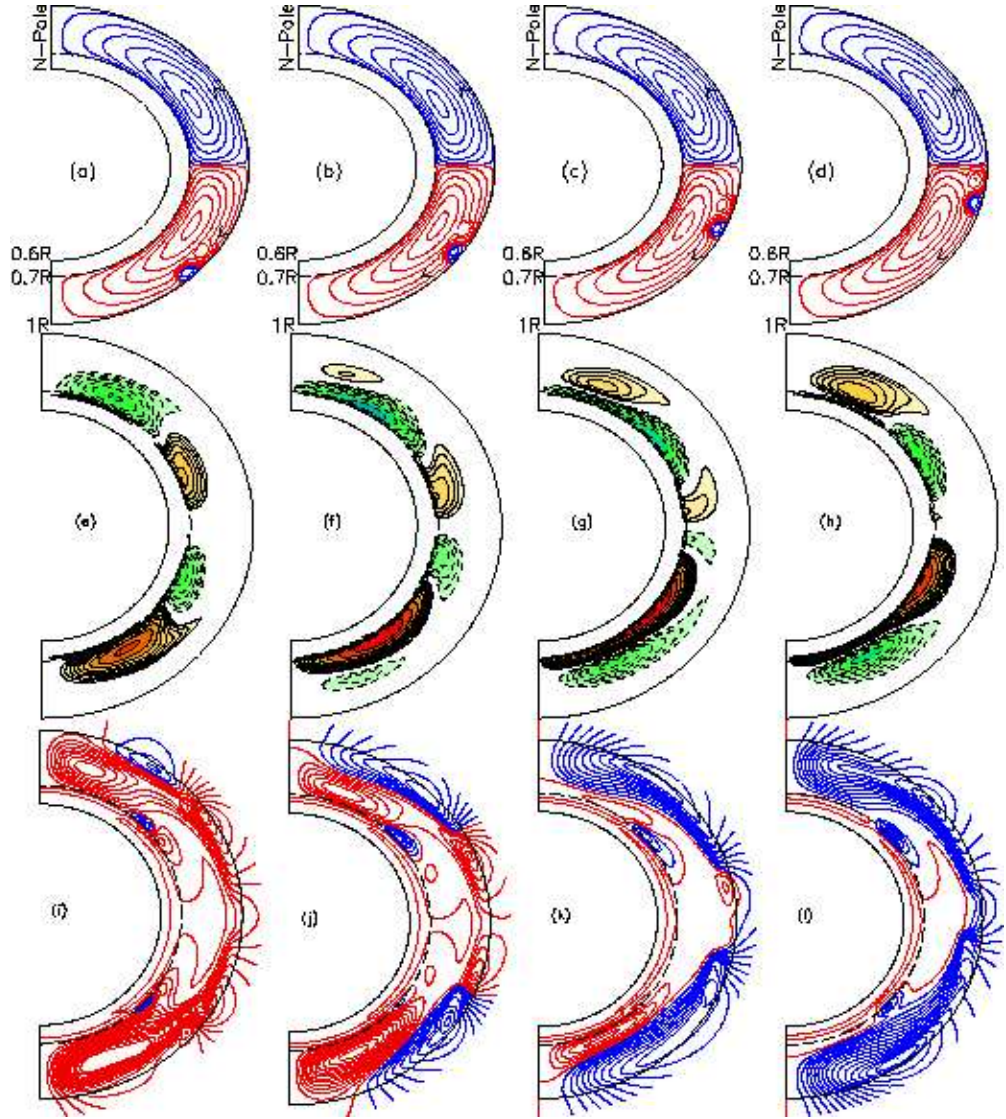}
\caption{Top panels (a-d): Snapshots of streamlines at four epochs within a sunspot cycle due to drifting of inflow cells from 
mid-latitude to the equator. Note that the South has stronger inflow cells in this simulation. Middle panels (e-h) show the 
snapshots of dynamo-generated toroidal fields at the same four epochs; red denotes positive field (going into the plane
of the paper) and blue negative field (coming out of the plane of the paper). Bottom panels (i-l) show the poloidal field lines; 
red denote positive (clockwise) and blue negative (anticlockwise).}\label{snapshots}
\end{figure}

In the simulations, the inflow pattern is introduced into both North and South hemispheres, but with a much stronger 
peak in the South (15 m s$^{-1}$ in the South versus 1.5 m s$^{-1}$ in the North). The choice of this difference is 
motivated by the fact that there were many more active regions in the South compared to the North in the declining 
phase of cycle 23 (see Fig.~\ref{decliningcycle23}) and the flux in the southern hemisphere was observed to be about a factor of 10 higher 
than in the northern hemisphere. In both hemispheres, the inflow circulation pattern is propagated toward the 
equator at a rate consistent with the equator-ward migration of the latitudes of active region appearance. 
Fig.~\ref{snapshots} shows the patterns of meridional circulation (panels a-d), toroidal field contours (panels e-h) 
and poloidal field lines (panels i-l) for a sequence of time intervals separated by 2.7 yr within a single sunspot
cycle. The simulation was begun a few cycles earlier with the same weak inflow in both hemispheres; the stronger inflow 
in the South was introduced a few months before the first frames shown. 

We see in panel (i) there is an immediate effect on the surface poloidal field 
in the South. By counting the number of contours of poloidal field lines,
we can see this effect in the form of more concentrated 
flux in the neighbourhood of the inflow pattern. Because the extra inflow near the surface is slowing down the migration of 
poloidal flux toward the pole, by panel (j) 2.7 years later, the polar field in the South is weaker and is reversing sign later 
than in the North. Again the number of contours reveals that
this results in less poloidal flux being transported to the bottom in high latitudes to cancel out the 
previous poloidal fields. This, in turn allows the toroidal field near the bottom in the South to become significantly stronger 
than in the North (see panels f,g,h). Therefore we see that the stronger inflows associated with one cycle in one hemisphere 
can lead to stronger toroidal fields in that hemisphere in the next cycle. This suggests that in the nearly independent North 
and South hemispheres the strength of one hemisphere compared to the other may persist for more than one cycle. There 
is observational evidence for this persistence, which is discussed in \citet[and reference therein]{DikGTG:2007} as well as differences in time of sunspot maximum. 


\clearpage

\begin{figure}[hbtp]
\centering
\includegraphics[width=0.8\textwidth]{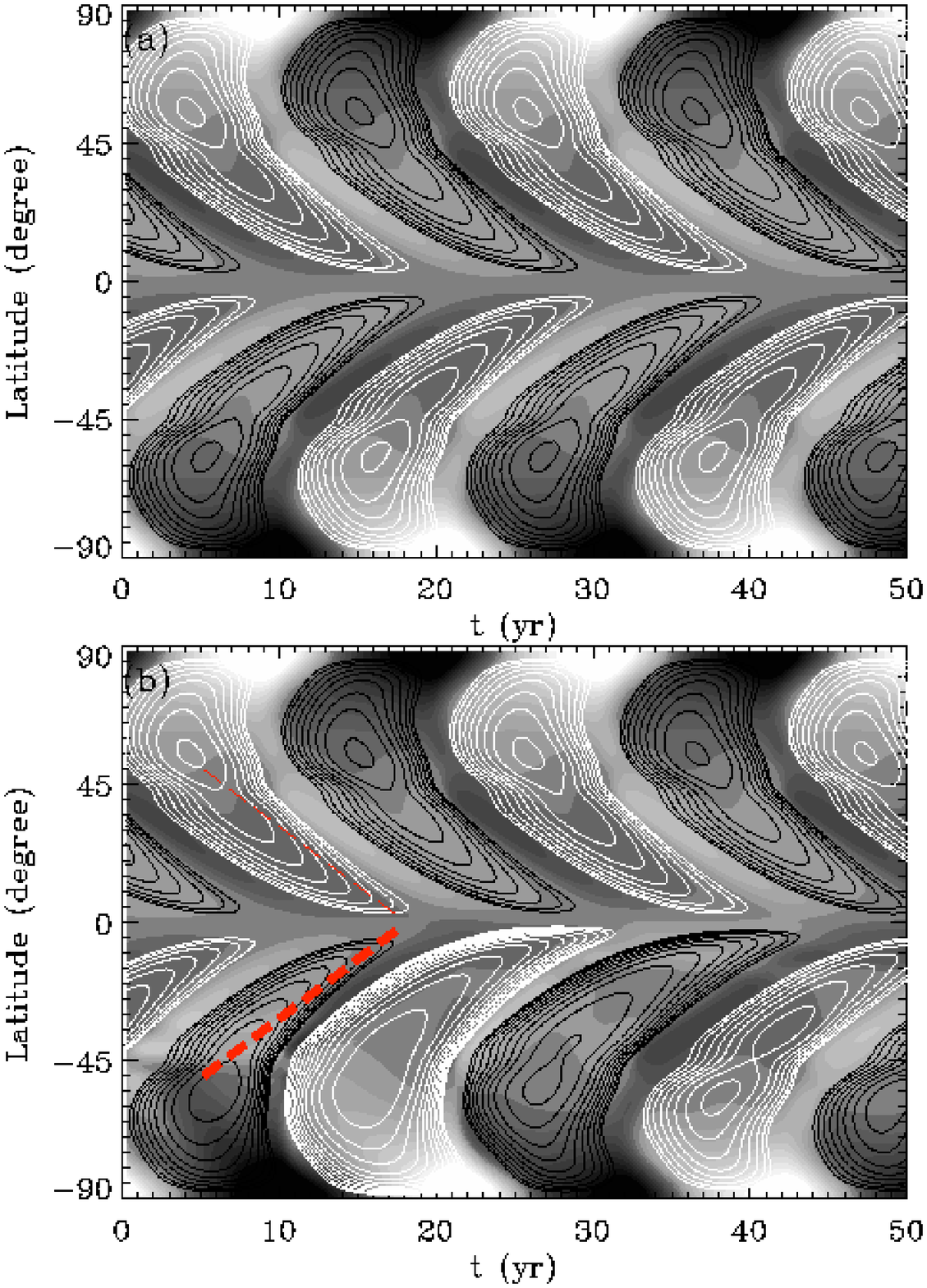}
\caption{Top panel (a): Time-latitude diagram of toroidal fields (black-white contours) taken from the base of the convection 
zone, and surface radial fields (greyscale map) for weak inflow cells of 1.5 m/s speed in both hemispheres. Bottom panel (b): 
Same as in (a) but for 10 times stronger inflow cell in the South compared to the North.
The locus of inflow cells are shown in red-dashed lines; a relatively thicker
line in the South denotes stronger inflow cells as incorporated in this 
simulation.}\label{butterfly}
\end{figure}

\clearpage

\begin{figure}[hbt]
\centering
\includegraphics[width=0.8\textwidth]{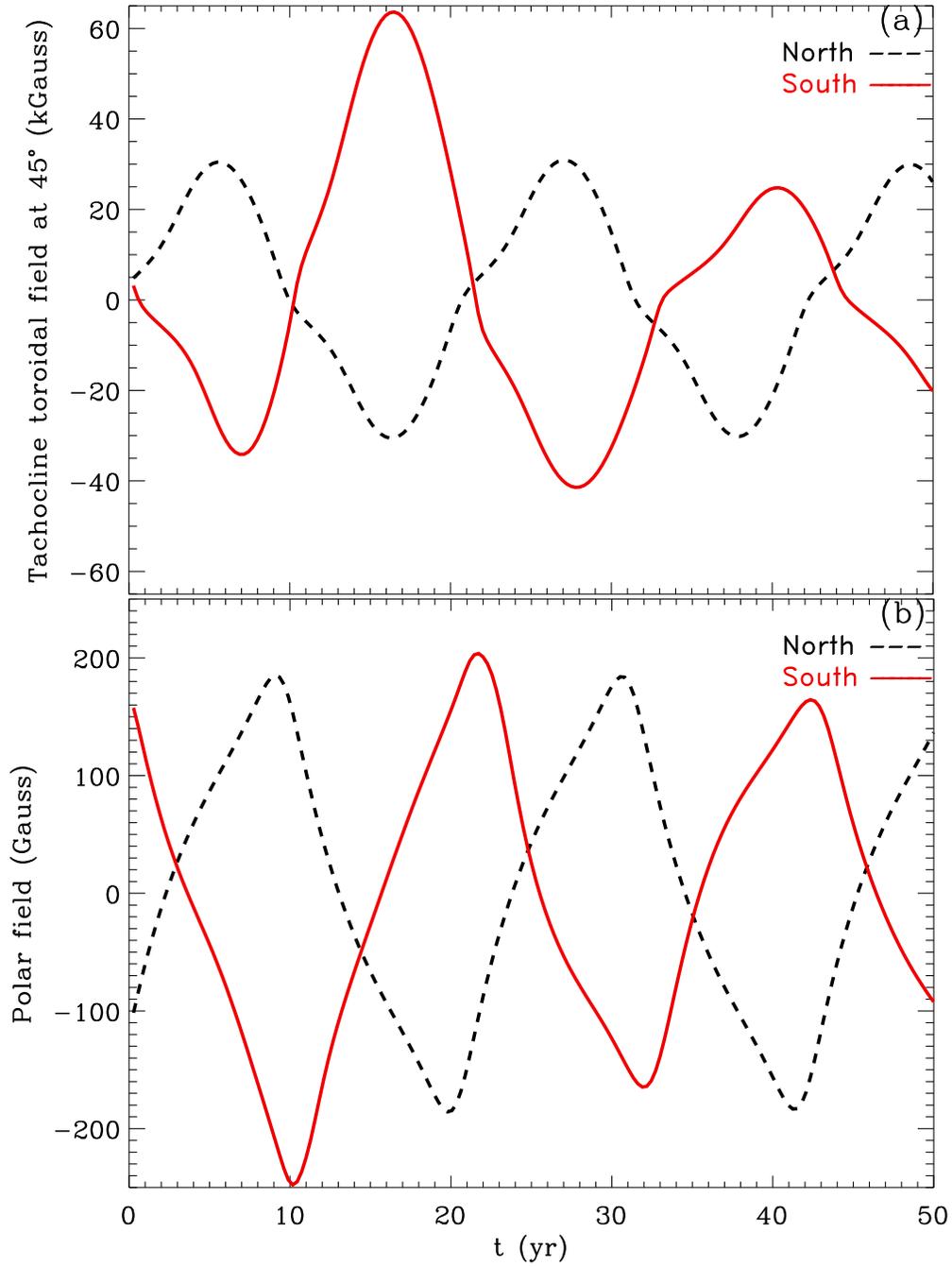}
\caption{Top image: Tachocline toroidal fields taken from $45^{\circ}$ latitude
in North (dashed black) and South (solid red) as function of time. Bottom image:
Polar fields in North (dashed black) and South (solid red).}
\label{toroidal-and-polar}
\end{figure}

\clearpage

In this particular simulation, the difference in inflow speed was introduced for a duration of about 12 years, after which the 
inflow in the South returned to the same lower value as in the North. Fig.~\ref{butterfly} shows a butterfly diagram for 
several cycles that includes the time with different inflows. On this diagram the extra inflow in the South occurred for years 
5-17. Shading is for the poloidal field amplitudes, contours for the toroidal field amplitude. If we focus on the toroidal field 
contours, we can see that the stronger toroidal field in the South persists for more than one cycle after the extra inflow has 
been switched off. This effect can clearly be seen in Fig.~\ref{toroidal-and-polar}, in
which the tachocline toroidal fields, taken from $45^{\circ}$ latitude, have been plotted
in the North (dashed black) and South (solid red) in the top frame. In the bottom frame 
the polar field patterns in the North (dashed black) and South (solid red)are presented.
The effect of even a temporary increase in inflow affects the dynamo well beyond the 
duration of the extra inflow; even though the maximum effect on the polar field, namely
a continuous increase in the South polar field can be seen during the drifting inflow 
cells until the sunspot minimum, the effect on the tachocline toroidal fields is 
more enhanced in the succeeding cycle, because of the increased polar fields being 
advected there by the time of the start of the next sunspot cycle, thus providing
a stronger seed magnetic field. \citet{Came2012} found a similar
effects, namely an increase in polar field at the end of a sunspot cycle and an
increase in the sunspot cycle strength in the succeeding cycle, due to the presence 
of inflow cells in their surface transport model. In reality the extra inflow would 
persist as long as more ARs are produced in the South, so the effect of this extra 
inflow would be even more pronounced and persistent, and inherently nonlinear.

\section{Summary and Discussion} \label{summary}

The peculiar behaviour of the solar cycle 23 and its prolonged minima has attracted much attention of the researchers over 
the last few years. There have been various studies taking very many different parameters into account. In the present 
paper we have discussed the contribution of AR's fluxes and their asymmetries in the north-south hemispheres, during the 
solar cycle 23 and rising phase of solar cycle 24 with the aim to address the issue of the deep minimum observed in the 
solar cycle 23. The observations showed that the cycle 23 was highly asymmetric. During the rise phase of the cycle 23, 
the northern hemisphere was dominant over the southern hemisphere which reversed during the decline phase of the cycle.

Further we concentrated our analysis on the declining phase of the cycle 23 and rising phase of the cycle 24. The analysis 
shows that the magnetic flux in the southern hemisphere is about 10 times stronger than that in the northern hemisphere 
during the declining phase of the solar cycle 23. The trend, however, reversed during the rising phase of the solar cycle 24 
and magnetic flux becomes more stronger (about a factor of 4) in the northern hemisphere. Moreover, it was found that 
there was significant delay (about 5 months) in changing the polarity in southern hemisphere in comparison with the 
northern hemisphere. These results may provide us with hints about how the toroidal fluxes would have contributed to 
the solar dynamo during the prolonged minima in the solar cycle 23 and in the rise phase of the solar cycle 24. 

It has been shown previously by \citet{2013ApJ...779....4B} that the degree of asymmetry in amplitude between North and 
South hemispheres can be changed significantly by differences in meridional circulation amplitude and/or profile between 
North and South. Here we have demonstrated that the difference between hemispheres in axisymmetric inflow into active 
region belts can lead to differences in peak amplitude that can last for more than one sunspot cycle. In the example shown 
here, we find that an increase in inflow in the South, which would accompany more solar activity there, leads to stronger 
toroidal fields in the South for substantially more than one cycle even after the extra inflow has been shut off. Therefore this 
mechanism can lead to persistence of one hemisphere dominating over the other for multiple cycles, as is often observed.
In effect, once a larger inflow is established in one hemisphere, its existence provides reinforcement for stronger cycles in 
that hemisphere to follow. An interesting question is then how the Sun eventually breaks out of this asymmetric pattern to 
a new one in which the other hemisphere dominates. Among other possibilities, this could occur when some other feature 
of meridional circulation, such as its amplitude or profile, changes in one hemisphere relative to the other.

\begin{acknowledgements} 
\textit{Acknowledgements:} 
We thank the referee for important inputs which has made the manuscript more comprehensive. Authors would like 
to thank Robert Cameron for valuable inputs on the manuscript. Juie Shetye 
thanks IUCAA for the excellent hospitality during her visit. This project was started at IUCAA. We gratefully acknowledge 
the Prof. John Gerard Doyle at Armagh Observatory for the important correction and suggestions. Research at Armagh 
Observatory is grant-aided by the Leverhulme Trust, Grant Ref: RPG-2013-014 and the work at High Altitude Observatory, 
National Center for Atmospheric Research Boulder is partially supported by NASA's LWS grant with award number 
NNX08AQ34G. The National Center for Atmospheric Research is sponsored by the National Science Foundation. SOHO is 
a project of international cooperation between ESA and NASA.\end{acknowledgements}

\bibliographystyle{apj}
\bibliography{references}
\end{document}